\documentclass[conference]{IEEEtran}
\IEEEoverridecommandlockouts
\usepackage{cite}
\usepackage{amsmath,amssymb,amsfonts}
\usepackage{algorithmic}
\usepackage{graphicx}
\usepackage{textcomp}
\usepackage{cite}
\usepackage[ruled,vlined]{algorithm2e}
\usepackage{romannum}
\usepackage{xcolor}
\def\BibTeX{{\rm B\kern-.05em{\sc i\kern-.025em b}\kern-.08em
    T\kern-.1667em\lower.7ex\hbox{E}\kern-.125emX}}

\newtheorem{lemma}{Lemma}

\IEEEoverridecommandlockouts\IEEEpubid{\makebox[\columnwidth]{ 978-1-6654-3540-6/22~\copyright~2022 IEEE \hfill} \hspace{\columnsep}\makebox[\columnwidth]{ }}
\begin{document}

\title{GRAND-assisted Optimal Modulation\thanks{The project or effort depicted was or is sponsored by the Defense Advanced Research Projects Agency under Grant number HR00112120008, the content of the information does not necessarily reflect the position or policy of the Government, and no official endorsement should be inferred.}}

\author{\IEEEauthorblockN{Basak Ozaydin}
\IEEEauthorblockA{\textit{Research Laboratory of Electronics} \\
\textit{MIT}\\
Cambridge, MA, USA \\
bozaydin@mit.edu}
\and
\IEEEauthorblockN{Muriel Médard}
\IEEEauthorblockA{\textit{Research Laboratory of Electronics} \\
\textit{MIT}\\
Cambridge, MA, USA \\
medard@mit.edu}
\and
\IEEEauthorblockN{Ken R. Duffy}
\IEEEauthorblockA{\textit{Hamilton Institute} \\
\textit{Maynooth University}\\
Ireland \\
ken.duffy@mu.ie}
}

\maketitle

\begin{abstract}
Optimal modulation (OM) schemes for Gaussian channels with peak and average power constraints are known to require nonuniform probability distributions over signal points, which presents practical challenges. An established way to map uniform binary sources to non-uniform symbol distributions is to assign a different number of bits to different constellation points. Doing so, however, means that erroneous demodulation at the receiver can lead to bit insertions or deletions that result in significant binary error propagation. 
In this paper, we introduce a light-weight variant of Guessing Random Additive Noise Decoding (GRAND) to resolve insertion and deletion errors at the receiver by using a simple padding scheme. Performance evaluation demonstrates that our approach results in an overall gain in demodulated bit-error-rate of over 2~dB Eb/N0 when compared to 128-Quadrature Amplitude Modulation (QAM). The GRAND-aided OM scheme outperforms coding with a low-density parity check code of the same average
rate as that induced by our simple padding. 

\end{abstract}

\section{Introduction}
\par In 1971 Smith proved that the optimal channel input of a scalar Gaussian channel under peak and average power constraints is a set of discrete points and that the optimal distribution over them can be determined by convex optimization \cite{smith}. Analogous results have since been established for other channels \cite{generalscalar, shamai, gursoy, poisson, conditionalGaus}. In the complex additive white Gaussian (CAWGN) channel with peak and average power constraints, it has been proved that the constellation points in the optimal channel input are discrete in amplitude and continuous in phase (DACP), forming concentric circles around the origin \cite{shamai}. Later, it was found that these continuous sets of points can be discretized with negligible loss in performance \cite{cons}. 
\par
\begin{figure}[t]
\centerline{\includegraphics[width=0.7\linewidth]{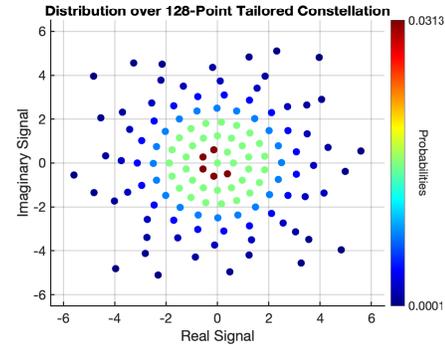}}
\caption{A non-uniform 128 point constellation identified in this paper, where the heatmap indicates the target symbol-use probability.}
\label{nonUnif}
\vspace{-0.15in}
\end{figure}

 
Assuming uniform binary sources, modulation schemes can be divided into two categories depending on the the frequency of their symbol use: uniform and non-uniform. Most commonly deployed schemes are uniform, including QAM and Phase Shift Keying (PSK), as used in applications including 5G, LTE, and IEEE 802.11 \cite{singya2020survey}. In contrast, Fig.~\ref{nonUnif} depicts a non-uniform modulation scheme that is designed according to the procedure that will be discussed in this paper. With the optimal channel input distributions of the CAWGN being nonuniform, studying the methods to identify nonuniform constellations is a worthwhile endeavour. As such, the aim of constellation shaping literature is to obtain enhanced power efficiency, which is achieved by nonuniform constellations.
\par Methods for identifying improved constellations include dividing constellation points into sublattices and using a binary error-correcting code on top of them \cite{voronoi,voronoi1,trellis}, or through varying the lengths of the bit sequences assigned to symbols to approximate target symbol-use distributions \cite{huff,Ungerboeck2002,dyadic,arith,ftv}. When the latter methods are used they are typically paired with an error correction code \cite{Ungerboeck2002}. For creating a variable length bit mapping, Huffman shaping is one technique that yields dyadic approximations to the desired target symbol-use probability distribution. In \cite{huff}, it is shown that this method can get very close to the maximum possible shaping gain of $1.53$ dB for unbounded Gaussian channels. Despite the core ideas underlying the design of optimal nonuniform constellations for peak-power constrained channels being known, they are not widely deployed. This is due to practical problems introduced by nonuniform distributions and the methods to identify them, which we discuss in the next paragraph.
\par In contrast to QAM or PSK where every symbol demaps to the same number of bits, the core practical challenge in using a modulation scheme where different symbols correspond to different numbers of bits is that it makes the demodulated binary sequence vulnerable to insertion and deletion errors. That is, if a symbol is erroneously demodulated to a symbol that corresponds to more or fewer bits, the demodulated bit sequence that follows will experience a shift that can result in significant error propagation. As a result, protection through coding is necessary. 
One approach is to add a binary forward error correction code after the symbol mapping\cite{ldpc,chain}. For instance, the authors of \cite{ldpc} use a LDPC code for this purpose. However, this approach comes at a cost of a large computational overhead at the receiver due to decoding. As for the nonuniform constellation design approach with sublattices, one needs to design a suitable code that would yield the channel optimal distribution. 
A core contribution of this paper is a new light-weight scheme based on the recently introduced Guessing Random Additive Noise Decoding (GRAND) 
\cite{GRAND,SGRAND,ORBGRAND} that enables the translation of improved symbol error rates to improved bit error rates through low complexity length correction based on symbol padding. While padding schemes have previously been proposed, e.g. \cite{Ungerboeck2002}, the one we employ is substantially simpler and is shown to be effective in protecting against insertion and deletions. 
\par In this paper, to realize OM: (\romannum{1}) we propose a new approximation for the cutting plane algorithm introduced in \cite{meyn} to facilitate the constellation design in the high signal-to-noise (SNR) regime; (\romannum{2}) we introduce a greedy algorithm for quantizing the continuous energy levels of the optimal channel input; (\romannum{3}) we design a simple padding scheme based on  Huffman shaping with low overhead and minimal complexity; and (\romannum{4}) we introduce a new light-weight GRAND variant that uses the padding information to correct the length of the transmitted message if an insertion or deletion event occurs.
\par Crucially, the padding scheme we propose to facilitate GRAND's operation is not a function of the data, resulting in significantly lower complexity operation than standard binary error correction coding schemes. 
With the proposed scheme, the padding overhead vanishes as SNR increases and the information rate approaches one. Indeed, in the proposed padding scheme the number of overhead symbols is significantly lower than with previously suggested schemes. Taking the padding overhead into account, simulation results show that, for a 128-point constellation, the system and constellation designs presented in this paper result in significant gains in symbol error rate (SER) and in bit error rate (BER) over QAM. 

\section{Channel Model and Notation}\label{cc}
Before detailing the system design, we introduce the paper's notation. 
Let $X$ and $Y$ be the complex channel input and output, respectively, and let $N$ be a complex Gaussian random variable, i.e. $N\sim\mathcal{N}_{\mathbf{C}}(0,N_0)$, with $N_0$ being the noise spectral density. The channel is modeled as 
    $Y = X + N$,
with $\mathbb{E}[|X|^2] \leq \sigma_P^2$ and $|X| \leq M$, where $M$ is the peak amplitude constraint and $\sigma_P^2$ is the average power constraint on the input distribution. $N$ is assumed to be independent of $X$, and its in-phase and quadrature components are distributed independently and identically. The channel capacity  is 
\begin{equation}
C = \max_{f_X(x)} I(X;Y)
\label{Cap}
\end{equation}
where the maximization is over all possible distributions for the channel input, i.e. $f_X(x)$, that satisfies the average and peak power constraints \cite{meyn}. We denote the set of distributions that satisfies those constraints by 
\begin{equation*}
{\cal M} = \{ f_X(x) : \mathbb{E}_{f_X(x)}[X^2] \leq \sigma_P^2\text{,}\; |X| \leq M \}. 
\end{equation*}
Let $X=A_Xe^{j\theta}$ and $Y=A_Ye^{j\gamma} $. 
The  conditional distribution of $A_Y$ given $A_X$ is given in \cite{shamai} as
\begin{equation}
\label{eqn:inoutAmp}
    f_{A_Y|A_X}(a_Y|a_X)=\exp\left(-\frac{(a_Y^2+a_X^2)}{N_0}\right)I_0\left(\frac{2a_Ya_X}{N_0}\right),
\end{equation}
where $I_0(x)$ is the modified Bessel function of the first kind and  $0^\text{th}$ order. With these definitions, the energy per symbol over noise spectral density is $E_S/N_0=\sigma_P^2/N_0$. For discrete modulation, we need to have a discrete set of possible values for $A_X$ and $\theta$, which we consider in the next section.

\section{Constellation Design}\label{T1}
There are two separate optimization problems when constructing close-to-optimal constellations. The first problem, which we consider in section \ref{AA} is finding the probability distribution of the amplitudes. A key result of \cite{meyn} is that an optimal $f_X(x)$ yields a distribution where the values $A_X$ or, equivalently, the energy levels, are in a discrete set, ${\cal A}$, while the phases are uniformly distributed.   Solving the first problem yields a DACP distribution. The second problem, which we consider in section \ref{BB}, is determining the number and the phases of the quantized points at each of the amplitudes found in section \ref{AA} in order to  yield a discrete constellation.
\subsection{Approximation to the Cutting Plane Algorithm}\label{AA}
A cutting plane algorithm to optimize the probability distribution over a fixed set of energy levels, $\mathcal{A}$, was previously proposed \cite{meyn} and shown to converge considerably more quickly than the Blahut-Arimoto algorithm. We provide a brief overview of that algorithm as applied to our setting. 
Reference \cite{meyn} solves \eqref{Cap} using a cutting plane algorithm that, instead of seeking directly the mutual information maximizing distribution $f_X(x)$, uses  a sequence
of increasingly tight relaxations, using approximations of the mutual information. 
 We define the channel sensitivity function $g \left( f_X(x), f_{X_i}(x) \right)$ as follows
\begin{equation*}
   g \left( f_X(x), f_{X_i}(x) \right) =  D\left( f_{Y_{f_X(x)}}(y) ||   f_{Y_{f_{X_i}(x)}}(y)
  \right), 
\end{equation*}
where $D( . || .)$ is the notation for the Kullback-Leibler divergence and $f_{Y_{f_X(x)}}(y) $ is the distribution of the output $Y$ when the input distribution is $f_X(x)$.
We begin with an initializing input distribution $f_{X_0}(x) \in {\cal M}$. At each iteration $n$ of the algorithm we solve the following  approximation to $I(X;Y)$
\begin{eqnarray*}
  & & I_n(f_X(x), f_{X_i}(x))  \nonumber  \\
 & = & \min_{0\leq i <n} E_{f_X(x)}\left[ g \left( f_X(x), f_{X_i}(x) \right)\right], f_X(x) \in {\cal M}.
\end{eqnarray*}
One can readily verify that $I_n(f_X(x), f_{X_i}(x))  \geq I(X;Y)$. Each iteration of the algorithm can be expressed as 
\begin{eqnarray}
    \max c  
    \text{ s.t.}  & E_{f_X(x)} \left[ I_n(f_X(x), f_{X_i}(x))   \right] \geq c, \nonumber \\
   &  0 \leq i < n,\quad f_X(x) \in {\cal M}.
   \label{optim}
\end{eqnarray}

\par To apply this algorithm to a CAWGN channel, we employ an essential modification to the  procedure that  \cite{meyn} uses. As $N_0$ decreases, the exponential term in $f_{A_Y|A_X}(a_Y|a_X)$, plummets whereas the modified Bessel function of the first kind increases steeply in \eqref{eqn:inoutAmp}, resulting in numerical issues for high SNR designs. To resolve this issue, we propose the following approximation for the Bessel function:
\begin{lemma}[\cite{bessel}]
\label{lem:bessel}
For large $|z|$,
\begin{equation}
\label{eqn:besselApprox}
    I_0(z)=\frac{e^z}{\sqrt{2\pi z}}\left(1+\mathcal{O}\left(z^{-2}\right)\right).
\end{equation}
\end{lemma}
Using this lemma $I_0(z)$ is replaced with $e^z/\sqrt{2\pi z}$. This provides simplifications when solving  (\ref{optim}), and removes a numerical indeterminacy in the calculation of $ g \left( f_X(x), f_{X_i}(x) \right)$.
\par For example, to obtain the 128-point constellation in Fig.~\ref{nonUnif}, the following parameters are used with this modified algorithm,
$\mathcal{A} = \{x|\;x = 0.6\,k,\; k=0,\dots,10\}$. The constellation design is made for $N_0=0.01$ with an average channel input power constraint of $\sigma_P^2=4$. The choice of $\sigma_P$ is the same as the parameter used in \cite{meyn} but the overall design SNR is chosen to be higher than the constellations presented there by taking $N_0=0.01$. The reason for this higher SNR choice is to target the values of engineering interest where the bit error rate of 128-QAM is around $10^{-4}$ in the design. In the results in the upcoming sections, we show that the proposed constellation design technique is robust and the constellations designed for $N_0=0.01$ also perform well at lower SNRs.
\subsection{Greedy Quantization} \label{BB}
While our discussion in \ref{AA} describes how to obtain ${\cal A}$,   the discrete set of amplitudes for the $f_X(x)$ that provides the maximization in (\ref{Cap}), the distribution is still continuous in phase. In order to have a true constellation and associated discrete input distribution $p_X(x)$, \cite{cons} shows how to discretize the inputs in a way that incurs a loss in capacity that vanishes with increasing $K$, the desired constellation size. The inputs at energy level $a\in\mathcal{A}$ can be represented by $k_a$ points, where
\begin{equation*}
    k_a = \left\lfloor\frac{\sqrt[\leftroot{-2}\uproot{2}3]{a^2p_{\mathcal{A}}(a)}}{\sum_{\Bar{a}\in\mathcal{A}}\sqrt[\leftroot{-2}\uproot{2}3]{\Bar{a}^2p_{\mathcal{A}}(\Bar{a})}}K\right\rfloor,
\end{equation*} 
where $p_{\mathcal{A}}(a)=\int_{|x|=a} f_X(x) dx$ is the total probability of the ring of amplitude $a$ in the DACP distribution found in \ref{AA}.
\par Each point $x$ on the energy level $a$ has the same probability, $p_x(x) = p_{\mathcal{A}}(a)/k_a$. After determining the number of points on the energy rings, the energy level $a\in\mathcal{A}$ is divided into $k_a$ arcs of the same length. To determine what phase offset each cycle should start with, a greedy algorithm is implemented. 
This algorithm aims to maximize the minimum distances between different energy levels. This is achieved by first arbitrarily placing every ring starting from the x-axis. Then, ring $i$ is rotated such that the minimum distance between the points on ring $i$ and the ring $i-1$ is maximized. 
Fig.~\ref{nonUnif} is a non-uniform 128-point constellation designed according to this procedure.
\section{Modulator Design}
\par The modulator design in this paper answers the question of how to realize the constellations discussed in section \ref{T1} by using a novel simple padding scheme with the Huffman shaping method, and a systematic way to perform varying-length bit mappings in an attempt to decrease bit errors. The proposed padding scheme is crucial for the GRAND-aided demodulator that will be introduced in section \ref{DM}.
\begin{figure*}[h]
  \centerline{\includegraphics[width=0.75\textwidth]{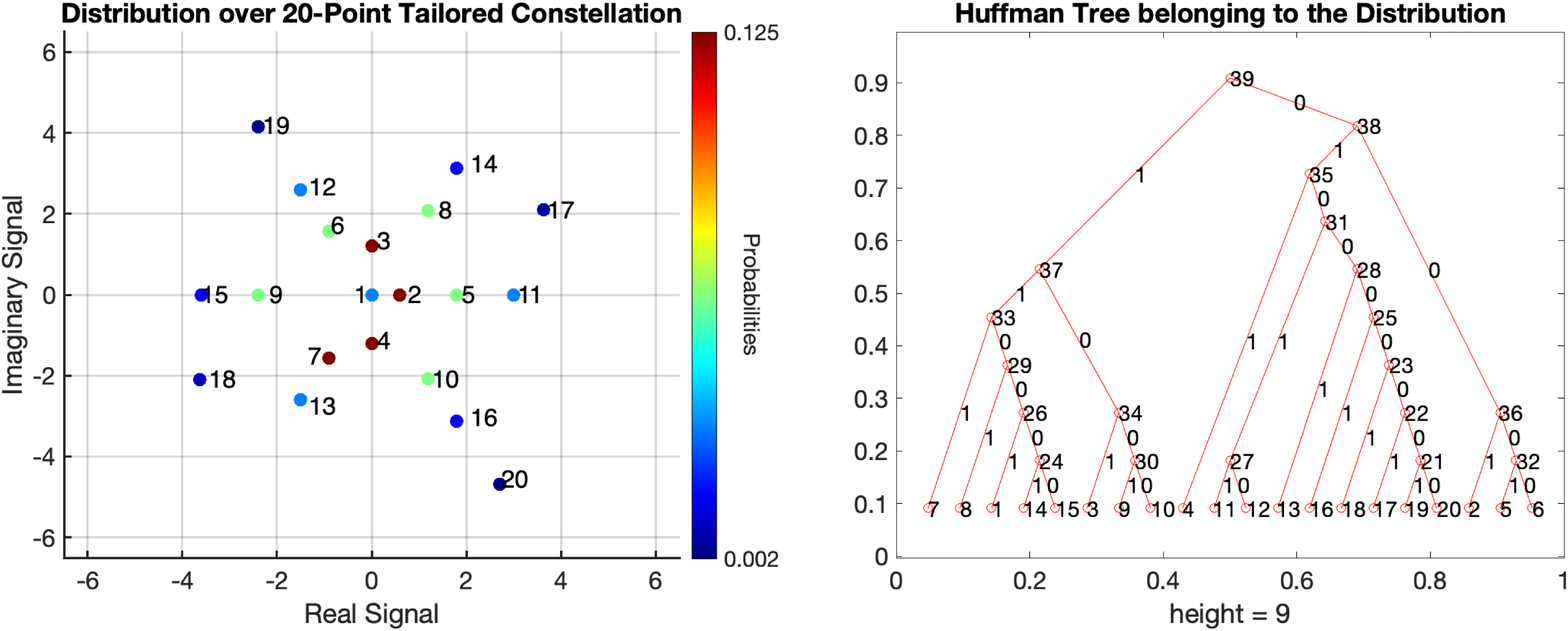}}
  \vspace{-0.1in}
  \caption{Illustrative 20-point constellation design and corresponding Huffman tree.}
  \vspace{-0.15in}
  \label{fig:huff}
\end{figure*}
In Huffman coding, symbols are encoded into bits using the paths from the root node to the leaf nodes of the optimal binary coding tree, whereas, Huffman shaping reverses this process by assigning symbols to bit sequences \cite{huff}. The optimal binary coding tree is not necessarily balanced, hence with fixed number of random bits as the modulator input, the path formed by the final untransmitted bits in the Huffman shaping method may not reach to a leaf node. In such cases, the following simple scheme, which GRAND will avail of for length correction, is applied. 
The input bit sequence is padded with a sequence of a ``1" and sufficiently many ``0" until the final bits are mapped to a symbol. 
If there is no bit in the original bit string that is not mapped to any symbol, padding starts at the root of the tree.
\par An illustration of this process is in Fig.~\ref{fig:huff}. If the bit sequence that arrived at the modulator is ``1110111", then the symbols 7 and 4 are transmitted according to Huffman shaping. The last bit does not reach a leaf and stops at node 37 of the tree. The proposed padding scheme dictates following the ``1" branch to node 33 and the ``0" branches are followed from there, leading to the transmission of symbol 15.
\par The proposed padding scheme is simple and effective when combined with the GRAND-based demodulator. The location of the last 1 indicates the end of the original message, hence the padding scheme provides the length of the original message. If the last received symbol is demodulated in error the location of the last 1 may change, but in section \ref{DM} we establish that by setting the input bit sequence length, $N$, appropriately, the padding frequency can be designed so that this happens sufficiently infrequently.
\par 
In most communication systems, a Gray code is used for mapping to constellation points that all have the same number of bits. We develop a different bit mapping algorithm, suited for OM, which identifies bit mappings of the same length in a greedy fashion such that the closest symbols that are represented by the same number of bits differ in only one bit. 
\section{Demodulation and Length Correction}\label{DM}
\begin{algorithm}[h]
\label{alg:lengthCorr}
\SetAlgoLined
\KwIn{$\mathbf{y}$: channel output\\
      \hspace{9mm} C: constellation points and their probabilities\\
      \hspace{9mm} f: symbol to bit mapping\\
      \hspace{9mm} $N_0$: noise power\\
      \hspace{9mm} n: number of transmitted bits}
\KwResult{$\hat{\mathbf{x}}$: demodulated bit sequence}
$\hat{\mathbf{y}}$ $\leftarrow$ MAP demodulation\\
$\hat{\mathbf{x}}$ $\leftarrow$ bit sequence corresponding to $\hat{y}$\\
\If{$\hat{x}$.length $\neq$ n}{
$i$ $\leftarrow$ 0, r $\leftarrow$ likelihood order of symbols in $\hat{\mathbf{y}}$\\
\While{$i<\hat{\mathbf{y}}.length$}{
$\mathcal{P}$ $\leftarrow$ set of constellation points on the same ring and the neighboring rings\\
p $\leftarrow$ likelihood order of elements in $\mathcal{P}$\\
\For{j= 1 to $\mathcal{P}$.length}{
    $\hat{\mathbf{x}}$ $\leftarrow$ bit sequence of $\hat{\mathbf{y}}[r(i)]=\mathcal{P}(p(j))$\\
    \uIf{$\hat{\mathbf{x}}$.length == n}{
        return $\hat{\mathbf{x}}$
    }
}
$i$ $\leftarrow$ $i+1$
}
}return bit sequence corresponding to $\hat{y}$
\caption{Demodulation and Length Correction}
\end{algorithm}
\par Unlike the case of uniform constellations, the demodulation process needs to incorporate the non-uniform prior distribution of the symbols, which can be achieved with Bayes’ theorem. As a result of the varying bit lengths of symbols, insertions and deletions can result from erroneous demodulation, which lead to bit error propagation and correspondingly avoid the large gains in the symbol error rate of using a non-uniform constellation from being reflected to the bit error rate. The modulation techniques that rely on these varying-length bit-to-symbol mappings attempt to remedy this issue through the use of coding techniques which inevitably yields a large computational overhead. Our novel GRAND-based demodulation corrects the bit sequence length by guessing the symbol that caused the length change and swapping it with the most probable length correcting symbol. Our demodulator therefore uses the padding as a form of an error correcting code, and can readily correct many of these length errors. The proposed procedure is presented in Algorithm~\ref{alg:lengthCorr}.
\begin{figure*}[t]
  \centerline{\includegraphics[width=0.65\textwidth]{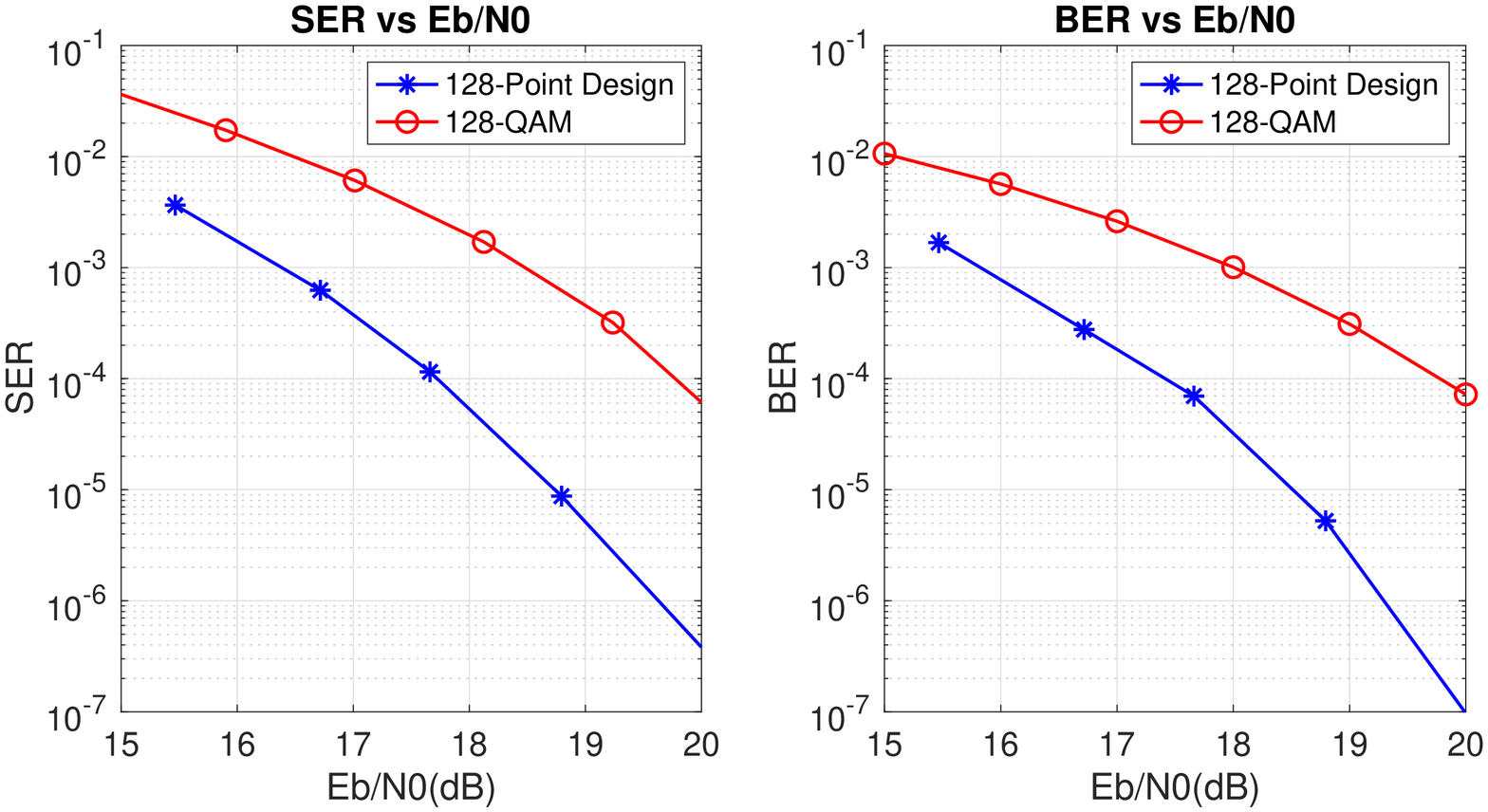}}
  \vspace{-0.1in}
  \caption{Symbol error rate (SER) and bit error rate (BER) performance of 128-Point design with GRAND-based length correction and 128-QAM.}
   \vspace{-0.15in}
  \label{fig:perf}
\end{figure*}
\par We use the term ``message" to indicate the information bit sequence and ``transmitted bit sequence" to include the padding bits. In the proposed system, the message length is constant and known by both the transmitter and receiver. However, the number of bits in the transmitted sequence changes based on the number of padding bits. Having removed the padding bits at the receiver, if the length of the remaining bits is not equal to the agreed message length, then at least one demodulated symbol resulted in an insertion or deletion.
\par 
Our goal in setting message lengths is to ensure that the likelihood of having more than one insertion or deletion error between the padded symbols is made negligible. For a given constellation and $E_S/N_0$, the probability of an insertion or deletion in demodulation, $p_{\text{indel}}(E_S/N_0)$, can be evaluated with theory or through simulation. According to the channel model presented in section \ref{cc}, symbol errors resulting in length changes occur independently. Hence we can express $N_l$, the number of symbols transmitted until a length changing error occurs is distributed geometrically with $p_{\text{indel}}(E_S/N_0)$.
Using this model, we need to set the message length, $N_s$, such that the probability of having more than one symbol error that results in an insertion or deletion within this sequence of $N_s$ symbols, $p_{N_s}$, is sufficiently small. To ensure this we impose the following constraint and solve the inequality for the largest $N_s$ satisfying it:
    $p_{N_s} \leq a(E_S/N_0)\,p_{\text{indel}}(E_S/N_0).$
In this inequality $a(E_S/N_0)$ is a small tunable parameter and we choose its exact value with a search to minimize the BER of the final binary system. All the results that will be presented in the next section are obtained after fixing the $a(E_S/N_0)$ values. 
Message lengths, $N_s$, for the reported simulations are presented in Table 1 for the 128-point constellation design in Fig.~\ref{nonUnif}. The message bit lengths are obtained by multiplying the number of symbols found via the above model with the weighted average length of the bit mappings.
\begin{table}[h]
\begin{center}
\caption{Some Message Lengths for 128-point Constellation Design}
\begin{tabular}{|c c c c|} 
 \hline
 $E_b/N_0\;(dB)$ & Ave. num.  bits ($N_b$)& Num. symbols ($N_s$)& Rate\\ [0.5ex] 
 \hline\hline
 20.00&1594&252&  0.998\\ 
 \hline
 18.75&771&122& 0.995\\
 \hline
 17.50&177&28& 0.979\\
 \hline
 16.25&56&9& 0.934\\
 \hline
\end{tabular}
\end{center}
\end{table}
\par In this design, message lengths increase with $E_s/N_0$, so that the rate of the overhead decreases with increasing SNR. As SNR goes to infinity, the overhead becomes infinitesimally small compared to the transmitted block length and the rate of the proposed scheme converges to one. In Table 1, the last column illustrates this change in rate for the 128 point constellation.
When an insertion or deletion does occur, it is most likely that there is a single erroneous symbol which results in shifting to a symbol on a neighboring ring, i.e. one energy level up or down. An ability to find the symbol that is in error and correcting it, or at least replacing it with one of the correct length, decreases both symbol and bit error rates. At this point, we draw inspiration from GRAND \cite{GRAND,SGRAND,ORBGRAND}.
\par First, the demodulator estimates the transmitted symbols via the decision regions given by Bayes' theorem and then converts the resulting symbols to a binary string. 
By construction, the bit sequence of the last symbol needs to contain a one. If there is none, then this indicates a false demodulation. An attempt can be made to correct the error by replacing the last symbol with the second most probable symbol for this signal.
\par The message length that is to be received is a constant known by the receiver. The demodulator compares the length of the message with this constant. If the lengths are not the same, then demodulated symbols are listed from least reliable to most reliable. Starting with the least reliable symbol, the demodulator examines the length of the bit sequence by switching the demodulated symbol to a symbol on its own ring or a neighboring ring. If there is a bit sequence of the correct length, then the original demodulated symbol is swapped with the most likely alternate symbol on these specified energy levels. If there is no bit sequence of the correct length, the demodulator proceeds to the next least reliable demodulated symbol. This process continues until a bit sequence of the correct length is found or all the demodulated symbols are exhausted. In the latter case, the original demodulated message bits are returned without any change. 
\par Consider the Huffman tree and the message bit sequence introduced in the earlier example. Suppose that after detection, the demodulator formed the received symbol sequence as $9-4-15$ which corresponds to the bit sequence ``1001011110000". After removing the padding, the remaining bits are ``10010111". The expected message length is 7, but the demodulator has 8 bits hence it attempts error correction. The received symbols are ordered according to their reliability. If symbol 9 was labelled as the least reliable symbol, the symbols on the same and the neighboring rings are listed as alternatives. According to Fig.~\ref{fig:huff}, the list in this example is $\{5,6,7,8,10,11,12,13\}$. The elements of this list are ordered according to their likelihoods and new bit sequences are formed by swapping symbol 9 with the elements of this list according to their likelihoods. When the demodulator finds a bit sequence with length 7, it terminates its search and outputs the resulting bit sequence. In this example swapping symbol 9 with symbol 7 yields the correct length, thus the demodulator outputs ``1110111" correctly.
\section{Results}
\par In this section, the performance of the 128-point OM constellation design with GRAND-like length correction is presented in comparison with 128-QAM, which is a standard in many applications. 
As the number of bits used to represent symbols may differ in the designed constellation, the conventional equation to calculate the $E_b/N_0$ does not hold. Define $N_s(E_s/N_0)$ as the average number of symbols for a message plus  padding bits. At most one symbol is used for the padding overhead. Hence, the average energy of a single information symbol is upper bounded by $E_s+E_s/N_s(E_s/N_0)$. As a result the following relation is used to obtain ${E_b}/{N_0}$:
\begin{equation*}
     \frac{E_s+E_s/N_s(E_s/N_0)}{N_0}= \frac{E_b}{N_0}\sum_{i=1}^K -p_i\log_2(p_i),
\end{equation*}
where $p_i$ is the probability of the constellation point $i$.
\begin{figure}[h]
    \centering
    \includegraphics[width=0.75\linewidth]{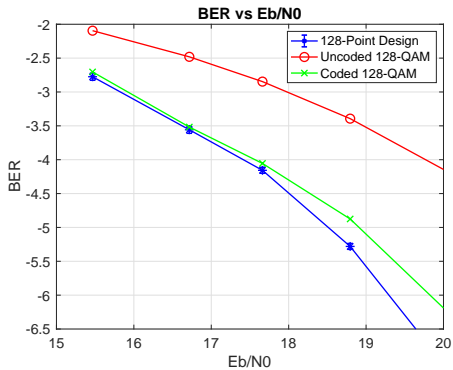}
    \caption{BER performance comparison of the 128-point design with uncoded 128 QAM and 128-QAM with an additional LDPC code for binary error correction. Due to the varying-length overhead in the proposed scheme, our 128-point design has different rates at different SNRs. The coded 128-QAM results at each SNR are obtained by using LDPC codes of the same average rate as the design at that particular SNR.}
    \label{fig:LDPC}
\end{figure}
\par In Fig.~\ref{fig:perf}, the SER and BER performance of the 128-point design equipped with message padding are presented with the blue starred line. For comparison, the performance of uncoded 128-QAM is shown by the red line with circles. While SER improvement is to be expected with non-uniform constellations, the additional error type of insertion and deletion makes it a challenge to translate that SER benefit to BER in the final binary system. Through light-weight padding and GRAND-assisted length correction, both of the plots in Fig.~\ref{fig:perf} show that the proposed scheme sees a gain of approximately $2$ dB over QAM. Crucially, this is achieved in a light-weight way and is transparent to the data receiver.
\par 
We note that $2$ dB is of the order that is typically attained through the use of computationally involved Forward Error Correction codes (FEC). To further assess that observation, we compared the BER performance of the proposed 128-point design and 128-QAM employing Low Density Parity Check (LDPC) codes. The choice of LDPC is due to the large block lengths, as seen in Table 1, we use in the high SNR regime and to given an estimate of where the proposed scheme may stand when compared to the conventional and wide-spread error correcting techniques. As the rate of the proposed scheme depends on the SNR, for the QAM with LDPC to be comparable with the proposed scheme, at each SNR a different LDPC code that has the same rate and the same block length as the proposed scheme is used. For instance, at $E_b/N_0=20$ dB, the LDPC code used in Fig.~\ref{fig:LDPC} has a message length of 1594 bits and a rate of 0.998. More message lengths and rates can be found on Table 1. We use the repeat-accumulate LDPC code design described in \cite{ld}. The decoder used with these LDPC codes is the built-in normalized min-sum LDPC decoder of MATLAB. The results displayed in Fig.~\ref{fig:LDPC} confirm the earlier finding that using OM with a simple padding scheme and GRAND-style length correction results in final BER performance that is as good as using computationally involved FEC schemes as an outside wrapper to standard modulation.
\section{Conclusion}
\par In this paper, we provide a system for making OM practical.
We present a design procedure to obtain non-uniform constellations according to channel statistics. It was already known in theory that optimal modulation schemes are non-uniform in their symbol transmission distributions, and that they can perform better in terms of capacity than the commonly used uniform modulation schemes. The proposed design provides a modulation and the associated demodulation schemes that can significantly surpass the performance of commonly used modulation schemes such as QAM. While OM is expected to be capable of providing significant SER benefits over commonly used schemes, to translate that to BER gains requires a method that can resolve insertion and deletion errors. Here we establish a simple, low-overhead, and computationally light mechanism to translate that gain to BER. Our method achieves this with a simple padding approach and a novel light-weight GRAND decoder, resulting in an improvement of the order of 2 dB that is transparent in the final binary data. The proposed system currently fixes one symbol when a length change occurs. Our future work includes extending our framework to fix more than one length-changing symbol errors and applying the proposed framework to different constellation sizes.


\bibliographystyle{unsrt}
\bibliography{cite.bib}
\end{document}